\documentclass{tmsce}

\usepackage{amsmath,amsfonts,amsthm}
\usepackage{algorithmic}
\usepackage{algorithm}
\usepackage{array}
\usepackage{booktabs} 
\usepackage[caption=false,font=normalsize,labelfont=sf,textfont=sf]{subfig}
\usepackage{textcomp}
\usepackage{stfloats}
\usepackage{url}
\usepackage{booktabs}
\usepackage{threeparttable}
\usepackage{verbatim}
\usepackage{graphicx}
\usepackage{siunitx} % Required for decimal alignment
\usepackage[style=apa, backend=biber]{biblatex}
\title{The Statistical Profitability of Social Media Sports Betting Influencers: Evidence from the Nigerian Market}

\authors{Kayode Makinde$^{1}$, Oluwatimileyin Onasanya$^{1}$, and Frances Adelakun$^{1,2}$}
\affiliation{$^{1}$ML Collective\\
$^{2}$Federal University of Agriculture, Abeokuta\\
\textit{Corresponding author:} makindekayode75@gmail.com}


\begin{document}
\maketitle

\begin{abstract}
This study examines whether following popular Nigerian sports betting influencers on social media is a financially sound strategy. To avoid the survivorship bias that occurs when influencers only share their winning bets, we tracked $5,467$ pre-match betting slips from three prominent tipsters on X (formerly Twitter) and Telegram. We verified the outcomes against official Stake.com records, resulting in a final dataset covering approximately $\$4.8$ million in tracked bets. We analyzed raw performance, assessed risk based on odds sizes, and applied four common staking strategies (Flat, Inverse, Square Root, and Fixed Return) to simulate realistic follower outcomes. The results show a sharp contrast between the wealth these influencers display online and the actual financial results. The influencers themselves collectively lost $25.24\%$ on their promoted bets, while a follower who staked the same amount on every tip would lose $38.27\%$ on their investment. Across all tested strategies, following these influencers consistently led to significant financial losses. These findings raise serious consumer protection concerns in Nigeria's expanding gambling market.
\end{abstract}
\printkeywords

% optional separator line (single-column)
\tmsceendfrontmatter

% =========================================================
\section{Introduction}
Deregulation and the spread of mobile technology have reshaped the global sports betting landscape. Once a niche activity, it has transformed into a major pillar of the entertainment economy. In North America, the 2018 Supreme Court decision to overturn the Professional and Amateur Sports Protection Act (PASPA) pushed sports betting into the cultural mainstream \parencite{robertson2024}. Africa has seen similar exponential growth; in South Africa, for instance, the betting sector grew by over $1,400\%$ in the five years leading up to 2023 \parencite{oluwole2025}, becoming a common form of household entertainment spending. This industry growth has also produced a new, highly influential, and largely unregulated byproduct: the sports betting influencer.

In Nigeria, widespread poverty and high usage of platforms like X (formerly Twitter) and Telegram have created ideal conditions for these influencers, some of whom command audiences in the millions. These individuals act as intermediaries between betting companies and everyday bettors. Instead of selling betting tips directly, many top influencers monetize their platforms through affiliate partnerships with sportsbooks \parencite{houghton2023}. Under this model, they receive commissions for bringing in new customers and encouraging continued betting volume. This setup creates a financial incentive structure that rewards ongoing user activity and recruitment, rather than the accuracy or profitability of the advice provided.

The financial mechanics of this arrangement are publicly available. According to \textcite{stake2026}, the affiliate commission formula for sportsbook activity is defined as:

\begin{equation}
\text{Affiliate Commission} = \left( \frac{0.03 \times \text{Wagered Amount}}{2} \right) \times \text{Commission Rate}
\end{equation}

Here, $0.03$ represents the platform's $3\%$ theoretical advantage, and the standard commission rate is $10\%$. Under this system, an influencer's earnings depend entirely on how much their followers bet, not on whether those bets win or lose. This creates strong incentives to keep followers engaged and betting, regardless of the accuracy of the predictions. This is particularly concerning given that the influencers in this study collectively command $4.7$ million followers.

We believe the luxurious lifestyles these influencers display, featuring expensive homes and cars, may lead followers to mistakenly believe their wealth comes from successful betting rather than affiliate payments. This carefully managed image exposes followers to severe survivorship bias. While a serious bettor relies on strategy to survive losing streaks, influencers heavily promote their ``roll-overs'' and ``guaranteed wins,'' sharing winning tickets prominently while ignoring evidence of losses. Although Nigeria's National Lottery Act of 2005 created a regulatory commission and prohibits issues like underage gambling and fraud, it does not require gambling to be marketed purely as entertainment rather than a viable career. This gap allows influencers to present sports betting as a realistic path out of poverty \parencite{nationalaws2005}.

Research from other countries suggests following online betting tipsters is financially dangerous. An observational study by \textcite{houghton2023} found that only about $20\%$ of ``expert'' picks promoted on Twitter actually won. Simulations of these advertised bets showed that following such advice consistently led to losses, with average capital depletion ranging from $12\%$ to $20\%$. Their study also highlighted selective result-sharing, noting that while affiliates publicized roughly $20\%$ of their wins, they acknowledged fewer than $2\%$ of their losses.

Despite these documented risks, no localized data exists measuring the actual financial impact of these influencers on Nigerians. This study aims to fill that gap by assessing the statistical profitability of popular African social media betting influencers. By tracking thousands of pre-match betting tips shared publicly before outcomes were known, we attempt to bridge the divide between the projected lifestyles of these influencers and the mathematical reality their followers experience.

\section{Methodology}
\subsection{Tipster Selection and Data Collection}
To evaluate sports betting influencers, this study focused on three prominent Nigerian tipsters: @mrbanks, @louiedi13, and @bossolamilekan1. These individuals were selected because they have the highest social media reach in Nigeria among influencers who frequently promote bet slips using Stake.com, boasting over 4.7 million followers combined on X and Telegram. Particularly notable is @mrbanks, who, with 1.9 million followers on X, is the most followed betting influencer on the entire platform globally \parencite{feedspot2026}.

To capture their betting activity without the survivorship bias that comes from self-reported wins, we built a data extraction system. We downloaded the complete message history from each influencer's Telegram channel, starting from when the channel was created up to the present day. The data was exported in JSON format. We then filtered the messages using keywords like "stake" and "bet" to find posts containing direct betting links.

\textit{It is worth noting, however, that one subject (@louiedi13) transitioned to a new channel in February 2026, but the $1,178$ tracked bets in this study reflect activity from the prior, higher-volume channel accessed during the primary data collection window (August--October 2025).}

We only included betting tips shared before the sporting events started. This pre-match requirement ensures that influencers cannot skew the data by only posting winning tickets after games end. For every valid bet link, we visited the Stake.com URL and recorded the verified bet details. For each bet, we collected: Message ID, Source (Tipster), Bet Link, Date Posted, Stake Amount, Odds, and Final Payout.

Table~\ref{tab:tipster_metrics} provides a breakdown of the total bets collected for each respective tipster during the two-year observation window. The collection period for the bets analyzed spanned over two years, from July 25, 2023, to August 24, 2025.

\begin{table}[ht]
\centering
\begin{threeparttable}
\caption{Breakdown of Social Media Presence and Betting Volume by Tipster}
\label{tab:tipster_metrics}
\begin{tabular}{lrrrr}
\toprule
\textbf{Tipster} & \textbf{X Followers} & \textbf{Telegram Subs} & \textbf{Tracked Games} & \textbf{Total Spend (\$)}\\ 
\midrule
@mrbanks         & 1.9M         & 466,787               & 3,677 & 4.31M \\ 
@louiedi13       & 1.0M         & 131\tnote{*}          & 1,178 & 0.32M \\ 
@bossolamilekan1 & 1.1M         & 209,628               & 612   & 0.16M \\ 
\midrule
\textbf{Total}   & \textbf{4.0M} & \textbf{676,546}     & \textbf{5,467} & \textbf{4.80M} \\ 
\bottomrule
\end{tabular}
\begin{tablenotes}
    \item[*] \small Subscriber count reflects a newly created channel as of February 2026. Data extraction and betting tracking for this subject were performed on a primary channel (since deleted or inaccessible) between August and October 2025.
\end{tablenotes}
\end{threeparttable}
\end{table}

\subsection{Platform Generalizability (The Stake.com Justification)}
Stake.com was utilized as the primary platform for this analysis because its infrastructure allows third-party users to publicly verify the exact stakes, odds, and payout statuses of shared bet slips via a direct URL. Most betting platforms operating in Nigeria obscure this information behind separate booking and verification codes, making large-scale, third-party verification impossible without placing the bets oneself. To ensure that the odds and payouts analyzed on Stake.com are a fair and generalizable representation of the broader Nigerian betting market, a comparative analysis was conducted using data from \textcite{oddsportal}. As shown in Table~\ref{tab:platform_comparison}, we compared Stake.com’s average payout rate against seven other leading bookmakers in the region: 1xBet, 22Bet, Cloudbet, BC.Game, Betking, Betano.ng, and Betway Africa NG.

The market average payout across these platforms was $92.95\%$. Payout refers to the proportion of total wagered money that a bookmaker returns to bettors through winnings (with the remainder representing the bookmaker’s margin). Stake.com demonstrated an average payout of $93.21\%$, ranking 4th out of the 8 bookmakers evaluated. Furthermore, Stake.com’s payout rate differed from the overall market average by only $+0.26$ percentage points, with an average absolute difference of just $0.49$ percentage points across all competitors. Because Stake.com’s odds align so closely with the median of the Nigerian market, the findings derived from this dataset can be confidently generalized to the broader sports betting ecosystem in Nigeria.

\begin{table}[h]
\centering
\caption{Comparative Analysis of Bookmaker Payout Rates vs. Stake.com}
\label{tab:platform_comparison}
\begin{tabular}{clSS}
\toprule
\textbf{Rank} & \textbf{Bookmaker} & {\textbf{Average Payout}} & {\textbf{Diff. from Stake.com}} \\ 
\midrule
1 & 1xBet            & 93.82\% & +0.61pp \\
2 & 22Bet            & 93.39\% & +0.18pp \\
3 & Cloudbet         & 93.31\% & +0.10pp \\
4 & Stake.com        & 93.21\% & 0.00pp  \\
5 & BC.Game          & 93.01\% & -0.20pp \\
6 & Betking          & 93.01\% & -0.20pp \\
7 & Betano.ng        & 92.53\% & -0.68pp \\
8 & Betway Africa NG & 91.28\% & -1.93pp \\
\bottomrule
\end{tabular}
\end{table}

\subsection{Data Preprocessing and Standardization}

Following the data extraction, preprocessing steps were undertaken to ensure consistency and accuracy in the financial analysis. The raw dataset contained bets placed in multiple currencies, specifically US Dollars (USD), Canadian Dollars (CAD), and Nigerian Naira (NGN). To facilitate a uniform analysis, all monetary values were converted into their USD equivalents using fixed exchange rates obtained in August 2025: USD (\$) at $1.0$, CAD at $0.72$, and NGN at $0.00065$.

Next, the dataset was scrubbed for administrative anomalies. Bets featuring odds of $\leq 1.00$ were strictly excluded from the performance analysis ($n=21$). These instances represent voided matches, canceled events, or partial refunds (e.g., Asian Handicap half-wins/losses) which are administrative sportsbook events rather than reflections of tipster predictive skill. These invalid entries represented less than $0.4\%$ of the total dataset.

After all cleaning, currency standardization, and filtering procedures were completed, the final dataset yielded $5,467$ valid pre-game bets. This comprehensive dataset represents a cumulative tracked betting volume of approximately $\$4.8$ million, forming the empirical foundation for the subsequent statistical and financial evaluations.

\subsection{Variable Definitions and Odds Categorization} \label{sec:odds_category}

To ensure rigorous evaluation, strict definitions were applied to the betting outcomes. In this study, a ``Win'' is exclusively defined as an event where the final payout amount is strictly greater than the initial staked amount. Any event resulting in a payout equal to or less than the initial stake (such as a partial cash-out at a loss) was recorded as a failure to yield a positive return.

Furthermore, to understand the risk appetite of the bettors and the relationship between expected returns and actual outcomes, the dataset was segmented into three distinct odds size categories. Because decimal odds represent the potential multiplier of the initial stake, these categories reflect the staker's expectations:

\begin{description}
    \item[Low Odds ($<10$):] Bets placed with the expectation of earning less than a tenfold return on the initial stake. These generally represent safer, lower-risk predictions.
    \item[Medium Odds ($10 \leq \text{Odds} \leq 100$):] Bets placed with the expectation of earning between a tenfold and a hundredfold return. This represents the vast majority of the dataset (accounting for approximately $50\%$ of the sample), indicating that the average follower seeks moderate-to-high returns.
    \item[High Odds ($>100$):] Bets placed with the expectation of earning more than a hundredfold return. These ``moonshot'' bets represent the highest risk category, where the statistical probability of the event occurring is exceptionally low.
\end{description}

\subsection{Evaluated Staking Strategies}
\label{sec:staking_strategy}
A main objective of this research is to determine whether systemic losses are a result of poor predictive accuracy by the tipsters, or simply poor money management. To investigate this, we simulated four distinct mathematical staking strategies against the dataset, drawing on foundational risk-management principles such as those outlined by \textcite{kelly1956} and \textcite{yao2006} to observe if different capital allocation methods could yield a profitable Return on Investment (ROI).

For the following formulas, let $S$ represent the calculated Stake amount, $O$ represent the decimal Odds of the bet, $C$ represent a constant base capital unit, and $P$ represent a target profit margin.

\begin{enumerate}
    \item \textbf{Flat Staking Strategy:} 
    This strategy assumes the bettor places an equal, fixed amount of money on every single bet, regardless of the odds or the perceived risk. It does not attempt to anticipate or adjust for the likelihood of an outcome. This approach serves as a standard baseline for assessing predictive performance, as described by \textcite{bargegil2020}.
    \begin{equation}
    S = C
    \end{equation}

    \item \textbf{Inverse Staking Strategy:} 
    A highly popular strategy among risk-conscious bettors, this method allocates capital inversely proportional to the total odds. It dictates staking large amounts of money on small odds (perceived high probability) and small amounts of money on high odds (perceived low probability). This proportional allocation of capital relative to the reciprocal of the odds was experimented on by \textcite{noon2012}.
    \begin{equation}
    S = \frac{C}{O}
    \end{equation}

    \item \textbf{Square Root Strategy:} 
    This strategy acts as a moderated, more efficient version of the Inverse Strategy. By taking the square root of the odds, it prevents the stake size from dropping too aggressively on high-odds bets, allowing the bettor to risk slightly more on long shots while still fundamentally favoring lower odds. This modification builds on existing square root staking principles, which are designed to control volatility and ensure gradual stake adjustment \parencite{marketfeeder2026}.
    \begin{equation}
    S = \frac{C}{\sqrt{O}}
    \end{equation}

    \item \textbf{Fixed Return (Target Profit) Strategy:} 
    Instead of basing the stake purely on capital preservation, this strategy works backward from a desired outcome. The bettor sets a fixed, predetermined profit margin ($P$) they wish to earn from every winning bet. The stake required for each bet is then dynamically calculated based on the odds to guarantee exactly that net profit if the bet wins.
    \begin{equation}
    S = \frac{P}{O - 1}
    \end{equation}
\end{enumerate}

By applying these four models to the $5,467$ pre-game predictions, the study evaluates not only the raw predictive power of the influencers but also the mathematical viability of their tips under optimal money management constraints.

\section{Results}

\subsection{Descriptive Statistics}

A total of $5,467$ verified pre-match bets were analyzed in this study. To understand the financial scale and typical behavior of the stakers, we examined three primary variables: the Stake Amount, the Odds Value, and the Payout Amount. Table~\ref{tab:summary_stats} summarizes these key metrics.

\begin{table}[h]
\centering
\caption{Summary Statistics of Tracked Bets (All monetary values in USD)}
\label{tab:summary_stats}
\begin{small}
\begin{tabular}{lrrrrr}
\toprule
\textbf{Variable} & \textbf{Count} & \textbf{Mean} & \textbf{Median} & \textbf{Min} & \textbf{Max} \\ 
\midrule
Stake Amount      & 5,467 & \$877.15    & \$300.03 & \$0.01 & \$80,348.71 \\
Odds Value        & 5,467 & 63,525.00   & 265.00   & 1.01   & 333,518,800.00 \\
Payout Amount     & 5,467 & \$655.72    & \$0.00   & \$0.00 & \$224,976.38 \\ 
\bottomrule
\end{tabular}
\end{small}
\end{table}

The average stake placed on a shared tip was \$877.15, yet the average payout returned was only \$655.72. The starkest indicator of risk is the median payout of \$0.00, confirming that the overwhelming majority of bets resulted in a total loss of capital.

The data is highly right-skewed across all metrics. For instance, while the median odds sit at $265$, indicating that the typical bettor hopes to win over two hundred times their initial stake; the maximum recorded odds reached astronomical, lottery-style figures (over $333$ million). Most importantly, across the $5,467$ tracked bets, the cumulative amount staked was approximately \$4.8 million, while the total returned payout was only \$3.6 million. This immediately highlights a massive, systemic depletion of capital within the observed betting pool.

\subsection{Win Rates and Tipster Profitability}

Further investigation into the outcomes of these $5,467$ bets reveals a harsh reality regarding the predictive accuracy of high-profile influencers.

\subsubsection{Overall and Individual Win Rates}

\begin{figure}[h]
    \centering
    \includegraphics[width=0.9\textwidth]{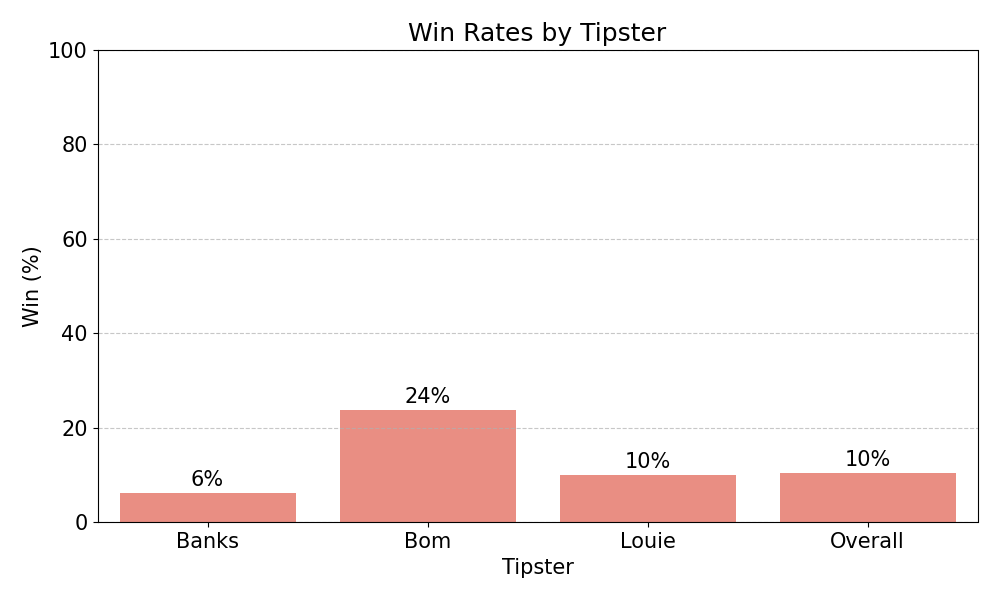} % Adjust filename as needed
    \caption{Win Rate across Tipsters.}
    \label{fig:win_rates}
\end{figure}

Across the entire dataset, the overall win rate was a mere $10.39\%$, with only $568$ successful bets out of the $5,467$ placed. A closer examination of the individual influencers, as shown in Figure~\ref{fig:win_rates}, reveals similar underperformance. Among the three tipsters, @mrbanks recorded the lowest win rate at just $6\%$ (accurately predicting roughly $1$ in $17$ games). @louiedi13 recorded a $10\%$ win rate, while @bossolamilekan1 (BOM) had the highest at $24\%$. However, as discussed in the following sections, a higher win rate does not inherently equate to financial profitability due to varying average odds sizes.

\subsubsection{Return on Investment (ROI) and Capital Loss}

To evaluate the tipsters’ true financial viability, we calculated their Capital Loss, which represents the negative of their Return on Investment (ROI). The ROI is defined as the percentage ratio of net profit to the total amount staked:

\begin{equation}
\text{ROI (\%)} = \left( \frac{\text{Total Payout} - \text{Total Stake}}{\text{Total Stake}} \right) \times 100
\end{equation}

Consequently, the Capital Loss is expressed as the negation of the ROI to illustrate the rate of wealth depletion:

\begin{equation}
\text{Capital Loss (\%)} = -\text{ROI (\%)}
\end{equation}

\begin{figure}[h]
    \centering
    \includegraphics[width=0.8\textwidth]{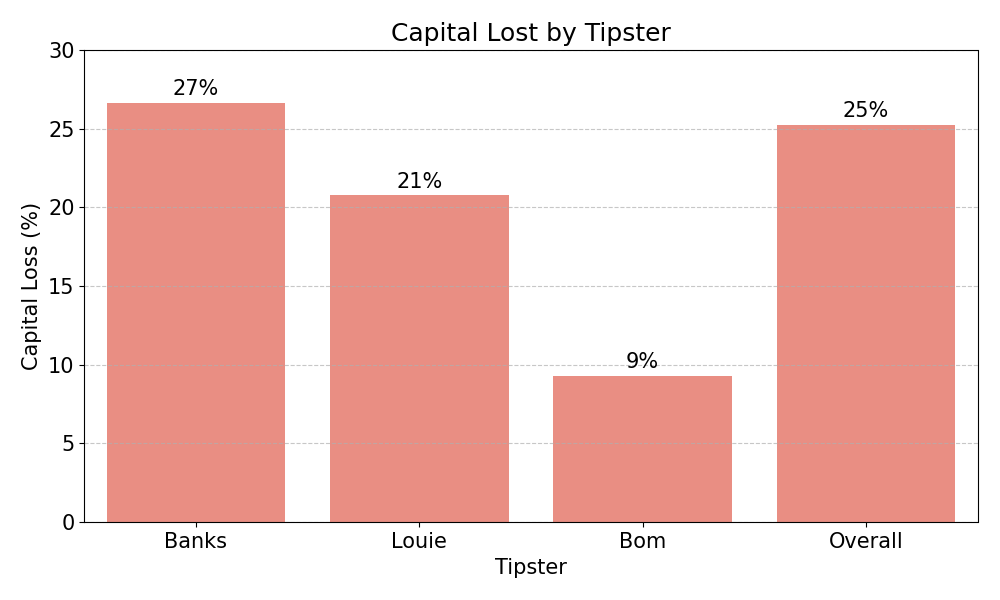}
    \caption{Comparative Capital Loss by Tipster.}
    \label{fig:capital_loss}
\end{figure}

A genuinely profitable tipster must generate a positive ROI (and therefore, a negative Capital Loss). In this study, all three influencers recorded substantial capital losses based on their own total staked amounts versus their payouts, illustrated in Figure~\ref{fig:capital_loss}. BOM demonstrated a $9.30\%$ capital loss, Louie recorded a $20.75\%$ loss, and Banks operated at a severe $26.60\%$ capital loss. Overall, following the exact stakes and tips of these influencers resulted in a total net loss of $25.24\%$ of all capital deployed. This is a stark contrast to the affluent, highly profitable lifestyles portrayed on their social media profiles.

% Note [h]: Add text labels to bars showing "Total Spend: $X / Total Payout: $Y"

\subsubsection{The Follower Simulation}

\begin{figure}[h]
    \centering
    \includegraphics[width=0.8\textwidth]{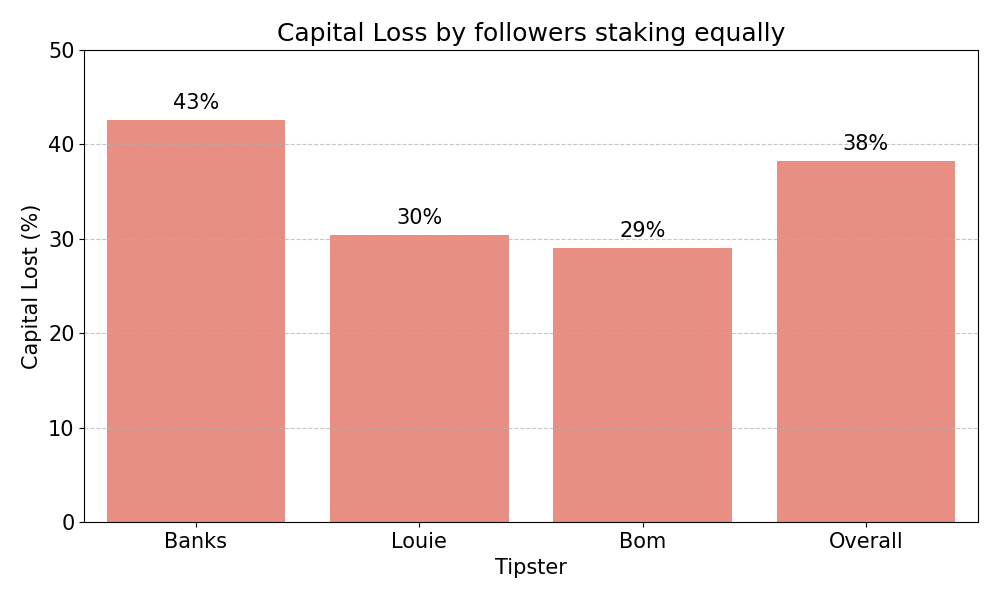}
    \caption{Capital Loss after Follower Simulation.}
    \label{fig:simulation}
\end{figure}

To quantify the real-world financial impact on an ordinary follower, we simulated a scenario where an average follower begins with an initial bankroll and blindly follows their chosen tipster using a flat staking approach. The outcome of this simulation is definitive: none of the tipsters generated a profit for their followers. As illustrated in Figure~\ref{fig:simulation}, followers would have depleted their initial capital severely, suffering absolute losses ranging from $29\%$  to $43\%$ depending on the tipster followed. This simulation indicates that adhering to these influencers' public predictions verbatim results in a high statistical probability of systemic capital erosion under the observed conditions.

\subsection{The Illusion of High Odds: Performance by Odds Size}

As defined in Section~\ref{sec:odds_category}, the dataset was categorized into Low ($<10$), Medium ($10$--$100$), and High ($>100$) odds to evaluate the risk appetite of the bettors and the viability of long-shot predictions. 

\begin{figure}[h]
    \centering
    \includegraphics[width=0.8\textwidth]{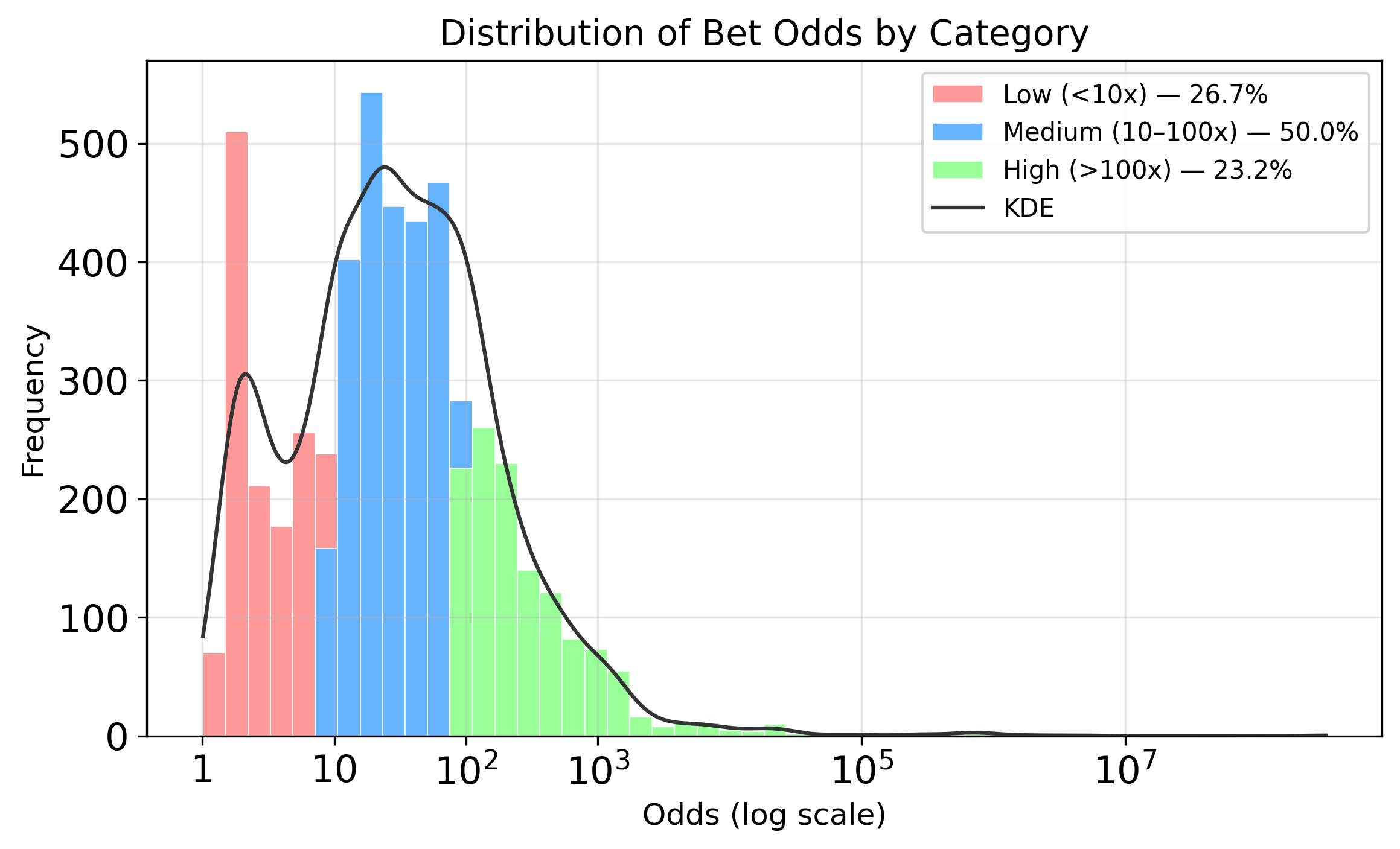} % Assuming Figure 4.6 is the distribution chart
    \caption{Distribution of Bets by Odds Category.}
    \label{fig:odds_dist}
\end{figure}

An analysis of the distribution reveals that the majority of tracked bets fell into the Medium odds category, accounting for approximately $50\%$ of the sample (see Figure~\ref{fig:odds_dist}). This indicates that the typical bettor follows tips with the expectation of earning between a tenfold and a hundredfold return on their initial stake. The Low odds category was the second most popular, comprising about $27\%$ of the bets, while the High odds ``moonshot'' category made up the minority.

\begin{figure}[h]
    \centering
    \includegraphics[width=0.8\textwidth]{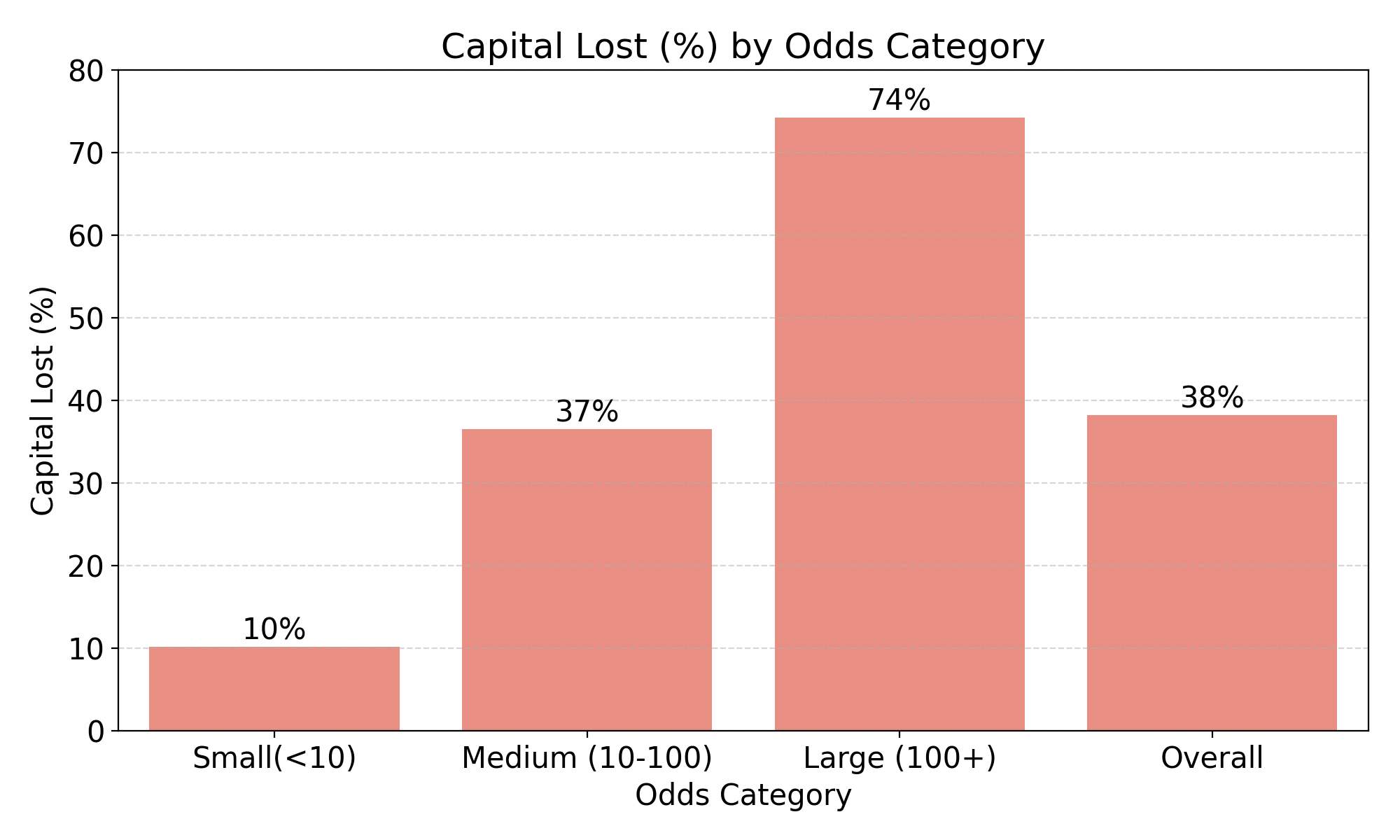}
    \caption{Comparative Capital Loss across Odds Categories.}
    \label{fig:odds_recovery}
\end{figure}

However, when comparing the actual financial recovery across these categories, a stark contrast emerges between bettor expectations and mathematical reality. As illustrated in Figure~\ref{fig:odds_recovery}, although the Medium odds category was the most heavily utilized, it consistently failed to yield long-term payouts exceeding the initial stakes.

The most ``favorable'' category was the Low odds segment ($<10$), which recorded a net loss of approximately $10\%$ of the initial stakes. This aligns with standard probability theory: bets with lower expected payouts carry lower variance and risk. Conversely, the High odds category ($>100$) proved to be statistically devastating. Followers chasing these massive payouts suffered an average loss of $74\%$ of their initial stake. This demonstrates that heavily promoted ``long-shot'' tickets carry a disproportionately high risk of capital destruction, and the matches included in these high-odds accumulators are highly unlikely to yield a win.

\subsection{Evaluation of Staking Strategies}
\label{sec:staking_strategy_results}
A rigorous investigation was conducted to determine if the systemic losses observed were strictly due to the tipsters' poor predictive accuracy, or if a disciplined money management system could extract a profit from their advice. We applied the four staking strategies defined in Section~\ref{sec:staking_strategy} (Flat, Inverse, Square Root, and Fixed Return) to the dataset.

\begin{figure}[h]
    \centering
    \includegraphics[width=0.8\textwidth]{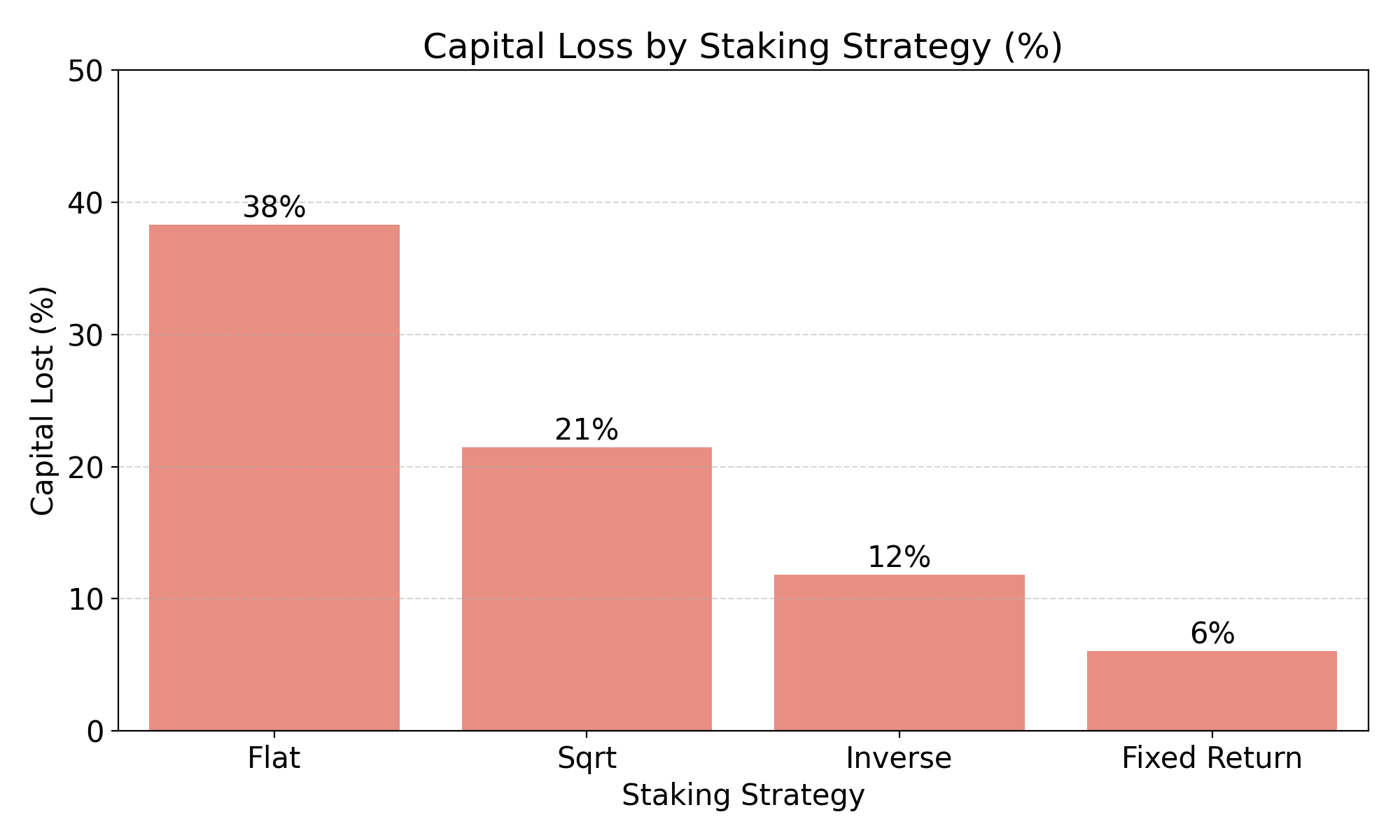}
    \caption{Capital Loss across Simulated Staking Strategies.}
    \label{fig:staking_results}
\end{figure}

The results, as shown in Figure~\ref{fig:staking_results}, show that while the choice of strategy affects the \textit{magnitude} of the loss, it cannot manufacture a profit. The Fixed Return strategy experienced the lowest marginal loss from the initial bankroll, acting as the most effective risk mitigation tool. This was followed by the Inverse Strategy, the Square Root Strategy, and finally, the Flat Strategy, which depleted capital the fastest. 

Most importantly, \textbf{none of the four strategies yielded a final bankroll greater than the initial capital.} This confirms that regardless of how carefully a follower manages their money, executing these influencers' betting tips results in an inevitable net loss.

\subsection{Statistical Inference Analysis}

To move beyond observational data and establish statistical certainty, One-Way Analysis of Variance (ANOVA) tests were conducted on the dataset.

\subsubsection{Effect of Staking Strategy and Tipster Choice on Profitability}

We conducted two separate ANOVA tests to evaluate the primary drivers of financial outcomes. The first tested the effect of the staking strategy, while the second examined the impact of the specific tipster followed.

\begin{itemize}
    \item[$H_0$:] There is no significant effect of the independent variable (Strategy or Tipster) on the betting results.
    \item[$H_a$:] There is a significant effect of the independent variable on the betting results.
\end{itemize}

\begin{table}[h]
\centering
\caption{Combined ANOVA Results: Impact of Strategy and Tipster on Profitability}
\label{tab:combined_anova}
\begin{small}
\begin{tabular}{lrrrc}
\toprule
\textbf{Source of Variation} & \textbf{Sum of Squares} & \textbf{df} & \textbf{F-Statistic} & \textbf{P-Value ($PR>F$)} \\ 
\midrule
\textit{Test 1: Staking Strategy} & & & & \\
C(Strategy)     & 10.890567               & 3           & 11.29327             & $2.126520 \times 10^{-7}$ \\
Residual        & 7028.120017             & 21,864      & ---                  & ---                       \\ 
\cmidrule(lr){1-5}
\textit{Test 2: Tipster (Source)} & & & & \\
C(Source)       & $9.253470 \times 10^{7}$  & 2           & 2.083428             & 0.124602                  \\
Residual        & $1.213408 \times 10^{11}$ & 5,464       & ---                  & ---                       \\ 
\bottomrule
\end{tabular}
\end{small}
\end{table}

With a p-value of $2.12 \times 10^{-7}$ ($p < 0.05$) for the staking strategy, we reject the null hypothesis for Test 1. There is a statistically significant difference in outcomes depending on the strategy used. However, contextualized with the findings in Section~\ref{sec:staking_strategy_results}, this simply means that structured staking strategies are effective tools for \textit{managing the rate of capital depletion}. They significantly alter how fast a bettor loses money, but their impact is practically limited, as even the mathematically optimized Fixed Return strategy failed to generate a positive ROI.

Conversely, the resulting p-value of $0.1246$ ($p > 0.05$) for the tipster source indicates that we fail to reject the null hypothesis for Test 2. Statistically, there is \textbf{no significant difference} between the financial outcomes of following any of the analyzed tipsters. While observational summaries may show superficial variations between @mrbanks, @louiedi13, and @bossolamilekan1, these variations are not large enough to rule out random chance. From a statistical standpoint, no single influencer in this study is mathematically superior; relying on any of them leads to the same outcome of systemic capital loss.

\section{Conclusion and Future Work}

\subsection{Conclusion}

This study set out to answer a critical socio-economic question: Does following the advice of high-profile Nigerian sports betting influencers make followers richer? By deploying a data-driven methodology to track $5,467$ pre-match predictions and completely bypass survivorship bias, the empirical evidence provides a definitive answer: No. Adhering to the public predictions of these influencers was consistently associated with systemic capital loss across all tested conditions in this study.

The findings reveal a notable disparity between the affluent lifestyles projected on social media and the documented financial outcomes of their tracked betting activity. Across an accumulated $\$4.8$ million in tracked wagers, the influencers themselves operated at a collective net loss of $25.24\%$. For the average follower, the reality is even more severe: systematically replicating these bets using a flat-stake approach yields a staggering $-38.27\%$ Return on Investment (ROI).

Crucially, rigorous statistical testing ($p = 0.1246$) confirmed that no single ``expert'' tipster performed significantly better than the rest. Furthermore, simulations of advanced money management systems including Inverse, Square Root, and Fixed Return staking proved that while structured staking alters the rate at which capital is depleted, no strategy was able to extract a positive return from the analyzed predictions. Ultimately, the observed disparity between projected success and documented betting performance suggests that affiliate marketing revenues may contribute significantly to these influencers' income, while the selective visibility of winning outcomes in their public posting history may further reinforce the perception of predictive profitability.

\subsection{Recommendations}

Given Nigeria's high poverty rates and the aggressive marketing of sports betting as a viable escape from financial hardship, the findings of this study demand immediate attention from both regulators and the public. To address these systemic risks, the National Lottery Regulatory Commission (NLRC), the body established to protect player interests under the National Lottery Act (2005), and other relevant bodies must implement strict disclosure mandates akin to those enforced by the \textcite{uk2026}, which binds all licensees to strict Licence Conditions and Codes of Practice (LCCP). Specifically, bookmakers and their affiliates must be legally required to explicitly market gambling solely as a form of entertainment, rather than as an investment strategy or a legitimate trade.

Furthermore, a framework for affiliate marketing transparency is required. Social media betting influencers should be mandated to visibly disclose their affiliate partnerships to their audience. Followers have a right to know when a tipster receives financial kickbacks for driving sign-ups or earns a percentage based on wagering volume. As this study has demonstrated, an influencer's income is often entirely independent of their predictive accuracy, creating a dangerous incentive structure that remains hidden from the casual bettor.

Finally, there is an urgent need for enhanced public financial literacy. Educational campaigns should actively highlight the impact of ``survivorship bias'' in social media betting. Bettors must be educated that high-odds accumulators (``moonshots'') carry catastrophic risk; destroying $74\%$ of capital, as evidenced in this study. By shifting the narrative from a path to wealth to a high-risk form of entertainment, the public can better protect themselves from systemic financial erosion.

\subsection{Limitations and Future Research}

While this study establishes a robust baseline using Stake.com for its verifiable public data, it does not capture the dynamic features of localized Nigerian platforms such as SportyBet or Bet9ja. Our simulations assumed static follower behavior, which does not account for user-specific actions such as early ``cash-outs'' or accumulator insurance (e.g., ``Cut-1''). Future research should employ an ``Active Bet Following'' methodology, placing minimum wagers on all shared tips across domestic platforms in real-time. This would validate our findings against localized odds margins and dynamic betting styles. Additionally, while standardizing multiple currencies (USD, CAD, NGN) into a single currency introduced minor distortions regarding absolute dollar totals due to Nigeria's exchange rate volatility, the percentage-based Return on Investment (ROI) calculations remain fundamentally sound.

Furthermore, the study's scope was limited to three mega-influencers operating on sportsbook affiliate models, leaving the profitability of smaller tipsters or subscription-based handicappers unexplored. Moving forward, research should expand the sample size across various influencer tiers. Future work must also directly quantify the extent of survivorship bias by tracking the silent deletion of losing predictions and analyzing how social media algorithms disproportionately amplify engagement on winning posts. Finally, to build upon these financial findings, qualitative survey research is needed to explore bettor psychology. Understanding why followers continue to trust mathematically unprofitable advice, whether driven by economic desperation or platform gamification, will be essential for developing effective financial literacy interventions.

\printbibliography

\end{document}